# FINAL REPORT
# WASTE MANAGEMENT

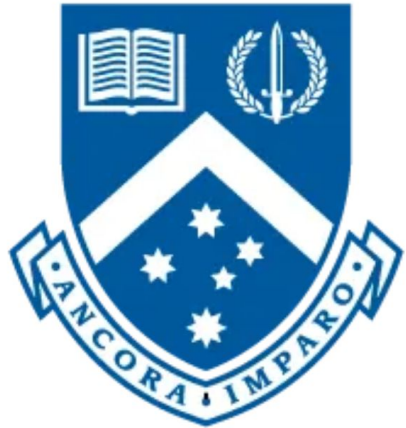

Team 16: An Tran, Fei Wu and Liqin Zhang

# Table of contents

# 1. Introduction: Brief intro to project and formulation of aims

Waste management is now the required feature of a modern city because professional waste management can bring up the health level of residents' lives of the city. Rubbish segregation is a crucial element in any sustainable waste management strategy for the healthcare section. Effective segregation of wastes means that there is less waste going to landfill which lets the cost on landfill become cheaper than without rubbish classification. Moreover, it is also better for protecting the environment and public health. Therefore, every facility should separate its waste at the source to reduce the cost of handling and disposal.

The goal of the project is to reduce cost and enhance the efficiency of waste management in the City of Melbourne. Teaching people how to classify rubbish is hard, but in an easier way, we can design a machine to tell them how to throw their rubbish in their hand. In addition, there are some moments that people do not know exactly which bin they should put the trash in. To solve it, we decided to create a mobile application so that all the residents can download it to their cell phone, the mobile app can easily estimate what is the item in front of the camera and tell the user which classification does it belong to, results in generalising the sense of rubbish classification to everyone lives in the city and finally controlling the rubbish segregation from the root part.

In this semester, all our team members worked on 2 main parts, building a machine learning model and developing a mobile application. We found a proper dataset and found a proper deep learning model to train the data, and then, we built an IOS application by using the best model we observed. Additionally, there were testing processes to check the performance of each part of the deliverables, tests on the accuracy of the deep learning model and tests on the final IOS application.

# 2. Background

## 2.1 Abstract of literature review

The main objective of our literature review is to assess different model implementations and methods to design a good model that can perform well on the TrashNer dataset. First of all, we want to understand the benefits of an intelligent system in enhancing waste management and waste planning. Then, we will have a look at the hybrid deep learning framework for waste classification. Finally, we want to understand how transfer learning can improve a model performance. In this part, we also discovers choice of lightweight models that can gives us as many advantages as it can on mobile devices

## 2.2 Introduction of literature review

The amount of municipal solid waste (MSW) generated is increasing rapidly due to urbanization and population growth, presenting unique opportunities and challenges (Korai et al). Solid waste is also widely known as "rubbish", the definition of solid wastes are useless solid, semisolid products left from consumption and production of humans and they are produced in all areas of society (EPA, south Australia). The solid wastes will be sent to various industries for recycling into another energy at global scale, to create jobs, reduce unemployment rate and increase the economy (Xu et al). In addition, MSW can be converted into different kinds of energy, MSW can be used as fuel in waste to energy plants, converting waste to electricity (Cucchiella et al). In our daily life, these solid wastes can be sorted into recycled waste, kitchen waste, Hazardous waste, and other wastes. Then these sorting useless solid wastes will be correctly recycled and converted into useful sources (Baidu, classification waste). Otherwise, unsorting waste solid wastes will increase load of industries, even using wrong processing approaches to these unsorting solid wastes can cause waste pollution to affect the physical environment and create aesthetic problems (Firdaus G.et al).

Therefore, our application can help individuals to correctly classify the waste and improves the efficiency of waste management bridge between individuals and waste industries.

## 2.3 Rational

In our project, we use deep learning CNN as our model. CNN model has many types of layers such as convolutional layer, pooling layer and fully connected layer. It can process lots of features of input which can easily handle the images as input with lots of pixels as features of input. CNN tries to simulate human vision which begins with eyes, but truly takes place in the brain. In human's vision, many biological neurons in the visual cortex react only to specific patterns in small regions of the visual field. As the visual signal makes its way through consecutive brain modules, neurons respond to more complex patterns in larger receptive fields. Like under picture, the images through the eyes and split information into different neurons. As the highest layer, it will group as fully information to our brain.

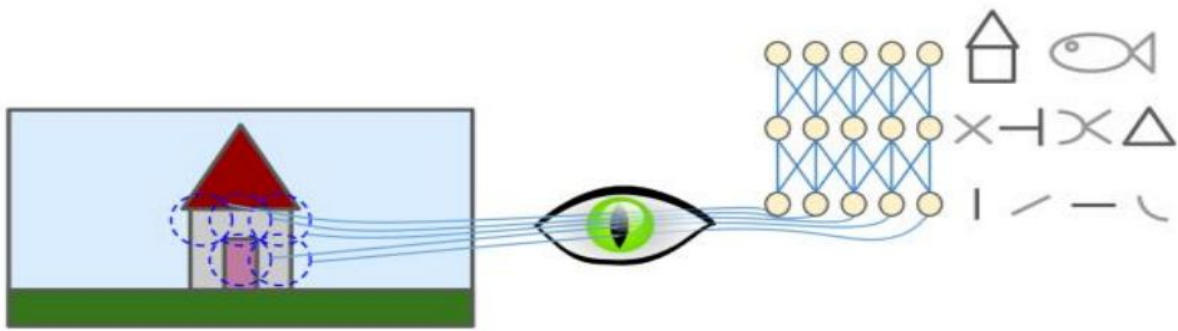

The CNN tries to simulate it, when the image goes through our model, the cells in different layers of the model try to split and process information and try to summarize it to send into a high layer. As the high layer, each cell will have rich information. Then it will be sent into a fully connected layer to classify.

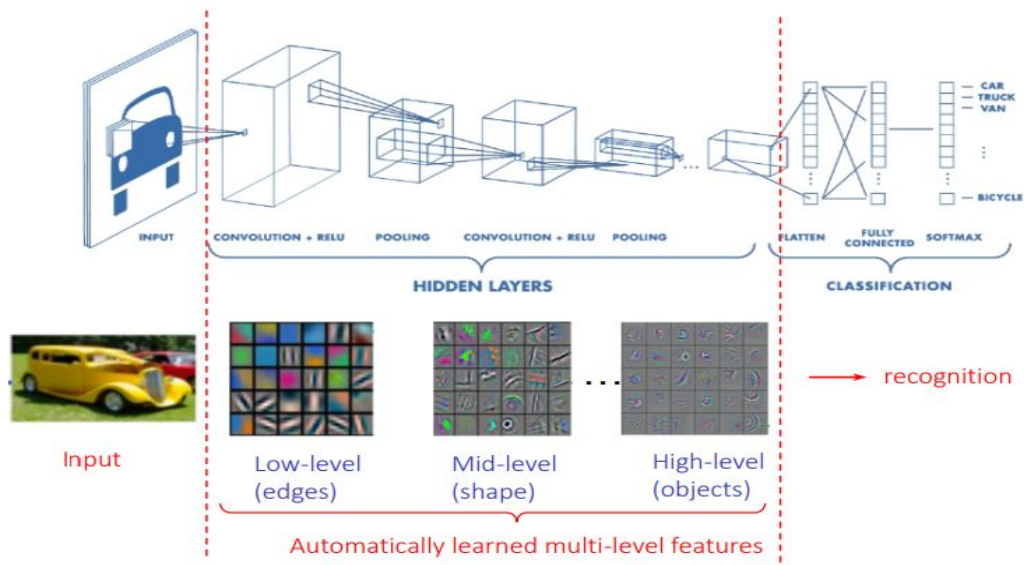

In addition, CNN can solve complex problems by changing it into simple questions. In other words, it will reduce high dimensionality into low dimensionality and changes high numbers of parameters into less numbers of parameters which can't affect the result, ie 1000*1000 pixels converted into 200*200 pixels does not affect the waste in the image. Unlike hand-crafted features which are manually engineered, the learnt features from machine learning algorithms like CNN reflect more details of the input and hence extract more insightful information. In the next part, we will discover how deep learning is used to build a classification model that can classify different types of trash images taken from the dataset. These research will be the building blocks for our projects as they give many insightful information on how to build a powerful model

## 2.4 Related research literature

We found 3 related researches about waste classification using machine learning.

**a. Application of deep learning object classifier to improve e-waste collection planning (Piotr Nowakowski, etc, al):**

This paper investigates a system to identify electrical and electronic equipment from images. The main purpose of this project is to improve the information exchanging from collection waste point and users. The system of this project was established on a mobile platform. Then, the mobile app will upload the images to the server, where the image will be processed through its server and classified. The methods of this project are using deep learning convolutional neural networks (CNN) to classify the e-waste and using a fast region-based CNN to detect the category and size of the waste from images. This paper shows that their accuracy can achieve 90%~97%.

In research, this network was using 180 images as a training data set. Each class has 60 images. For the test, it uses 30 images and 10 for each class. Each image will have one waste item like refrigerator, washing machine or monitor (three classes). Each image was scaled to 128 *128 pixels. For CNN, the images of e-waste go through the CNN classifier. The machine learning algorithm uses computational methods to learn information from data without relying on a predetermined equation as a model. To recognize several objects in one image, faster R-CNN was used. Actually, CNN was used as the basis of faster R-CNN detection. The R-CNN gets the result from CNN to detect and give labels to each image, which will increase the accuracy of this project.

In summary, this project improves the efficiency of collection points and individuals to exchange waste information. This new method for identifying waste enables the development of a ready-made digital solution to recognize the equipment reported for collection based on customer images. However, it also has some shortcomings. The numbers of classes for this project is not enough. It only has three kinds of e-waste and can't identify the other e-waste like laptops or table lamps. In addition, less numbers of data sets can't get a better model. Some images of broken e-waste are not contained in this project. This also can challenge this project.

**b. Multilayer Hybrid Deep-Learning Method for Waste Classification and Recycling (Yinghao Chu et al)**

The objective of this study is to use a multilayer hybrid deep learning system (MHS) to classify waste automatically. This project needs two kinds of hardwares, one of them is a high- resolution camera to get a waste image and the other one is a sensor to extract other useful features like weight. In this paper, the project will use MHS to identify the waste based on images whether it is recyclable or not. For algorithms, the MHS used CNN as base algorithm for capturing image features and multilayer perceptrons (MLP) to enhance accuracy of this model. In other words, based on general layer distribution of CNN, this project adds some more layers and some active functions as multilayer perceptrons.

This MHS model trained 5000 images and tried to classify them into recyclable waste kinds—paper, plastic, metal and glass. The other kind as one class contains fruit, vegetable and plant. Compared with a single CNN model with 80% accuracy, MHS achieves accuracy rates above 90%. Therefore, this study proves their MHS system can play a good performance to sort waste into recycle kind and other kinds.

In summary, this study addresses the problems of house waste. The MHS can identify recycled waste with high accuracy and improve the house waste management.  In considering the continually increased volume of waste globally and the urgent requirements for environmentally friendly waste processing, their MHS system is both economically and environmentally beneficial. Moreover, this study also shows us that the MHS system can play better performance than traditional CNN classification. CNN is one of the most recognized deep learning algorithms for wide application in image classification, segmentation, and detection (A. Krizhevsky,et,al)( K. He, X. Zhang,et,al). The MLP as a new structure of classification can be used into more fields. However, it has some shortcomings in this study. There are only two kinds for identification, the one is recyclable and the other one is others. But, for industry, they have at least 4 kinds of solid waste--recycled waste, kitchen waste, Hazardous waste, and other wastes. So this project can't meet the requirement of industry to classify waste.

**c. Reach On Waste Classification and Identification by Transfer Learning and Lightweight Neural Network (Xiujie Xu, Xuehai Qi, etc)**

This project presents a waste classification and identification method using transfer learning and lightweight neural network. By migrating the lightweight neural network MobileNetV2 and rebuilding it, the features will be introduced into the SVM to classify 6 kinds of garbage. In addition, its model was trained by using 2527 images of waste labeled in TrashNet dataset which resulted in classification accuracy 98.4%. It means the methods of this project can improve classification accuracy and overcome the problem of weak data and less labeling. For deep learning of this project, it uses genuinely convolutional neural networks (CNN). The structure of CNN consists of convolutional layer, pooling layer and fully connected layer. In addition, the layer distribution in this project is following the VGG network, then the performance of its model can be improved by increasing the depth of the network. Moreover, this project uses the transfer learning methodology and lightweight network created by Google (MobileNetV2). It uses this network to enhance and increase its project accuracy, based on the VGG network.

Traditional machine learning and deep learning to solve classification is efficient. However, the generation of massive data theoretically makes machine learning or deep learning have objects to rely on, but these data are often in primitive form and manually labeling the data is time-consuming and labor-intensive. This project using transfer learning and lightweight to identify waste by images only uses less numbers of images and less computation time to get a good model.

## 3. Methodology

### 3.1 Methodology implementation

Our dataset used in our project is from TrashNet, collected by Thung G et al. In the TrashNet dataset, it has 2527 images of waste including 501 glass images, 594 paper images, 403 cardboard images, 482 plastic images, 410 metal images and 137 trash images.

We used this dataset to implement a convolutional neural network (CNN) using Tensorflow. Tensorflow is an end-to-end open source framework for machine learning and deep learning developed by Google that provides high-level APIs. On top of Tensorflow, we used Keras as a main library as it allows fast and easy prototyping of a network run smoothly on both CPU and GPU. The main platform for model creation is Google Colab as it provides a free GPU which helps us accelerate the training process

We implemented a total of three models: 2 models built from scratch and 1 built using transfer learning technique. To achieve this, we first created a common pipeline to import and process images from our root folder. We defined an ImageDataGenerator variable using Keras and used its flow_from_directory method to load all images from the folder. This particular implementation will allow us to augment data in real-time at ease, hence increasing the number of data in the dataset. Then, we build our models and simply fit the dataset into our models. We picked the best model and converted it into Tensorflow Lite format before deploying it into our application.

The application is an iOS mobile application. This application has four main modules, which takes different responsibilities in the app. We will discuss the architecture of the app in more detail in the design part. In brief, we implemented a module called ModelDataHandler to receive and process input as well as return the final result to the View of our application. The input will be the frames of the picture captured by a camera manager module. The app does inference by leveraging the TensorflowLite library for Swift also built by Google.  In addition to MVC, there is also a dedicated module to manage camera called CameraFeedManager. This class will contain all the functionalities related to camera.

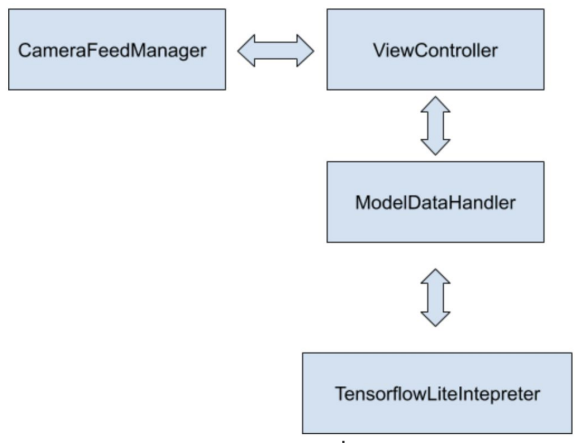

Tensorflow Lite also supports hardware accelerators on iOS so we used GPU delegate for inference. As discussed above, GPUs are optimized to advance deep learning model training because GPU can handle multiple computation simultaneously. Moreover, GPUs can eliminate the concern of decreased accuracy on quantization optimization due to strong computation ability with both 16-bit and 32-bit floating point. Finally, GPU optimized computation can help reduce energy consumption for the same task compared with CPU.

Moreover, we also allow the app to change the number of threads used for classification tasks. However, it depends on the type of task to modify the suitable number of threads as there is a trade off for latency and energy efficiency when we increase the number of threads.

## 3.2 Design

In this section, we will discuss the architecture of our models and then the architecture of our application implemented in the previous part.

The first scratch CNN is built by a block of the pattern [conv, batch norm, activation, conv, batch norm, activation, mean pool, drop out] and then attached a flatten layer and a dense layer with softmax to produce final results that match the number of classes in the database.

The second model is built with VGG16 architecture proposed by K. Simonyan and A. Zisserman in their paper "Very Deep Convolutional Networks for Large-Scale Image Recognition". We used this model because this architecture is one of the best models that performs very well on ImageNet, a dataset with over 14 million images belonging to 22,000 categories. As a result, we believe this architecture will help us achieve adequate results for similar tasks. VGG16 consists of 16 layers with uniform architecture. They are a stack of convolutional layers with the smallest filter size of 3x3 followed by a max pooling layer of stride 2x2. The last 3 layers are fully-connected layers with the output matching the number of classes. VGG 16 has more than 138 million parameters to train. In our model, we will modify the suitable output for last dense layer.x

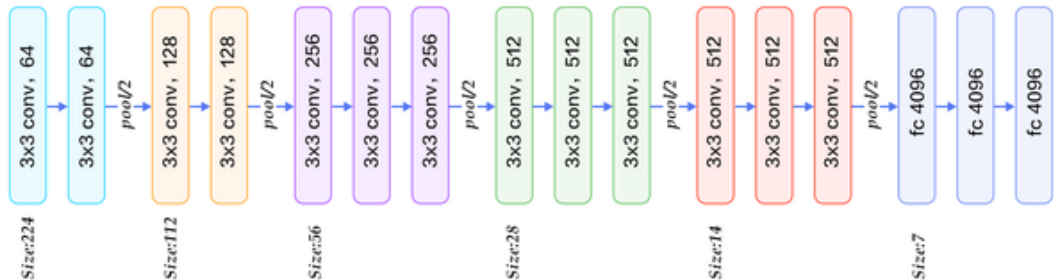

The last model is built by transfer learning techniques. This model has the base of MobileNetV2 model. This model is a light weight model that seeks to perform well on edge devices like mobile phones. This model is built by Google based on the previous idea of depthwise separable convolution introduced on MobileNetV1 coming with a novel module called inverted residual structure. The architecture of MobileNetV2 contains the convolution layers followed by 19 residual bottleneck layers and a few more convolution layers to gain output. The primary network of MobileNetV2 uses around 2.2 million parameters. However, these parameters are not trainable in our model. On top of the MobileNetV2, we can simply implement a dense layer to match the number of classes we have. However, in our model, we added two blocks of [dense, batch norm, drop out] to maximise the performance of this model. The additional parameters from these two blocks are more than 5 million parameters.

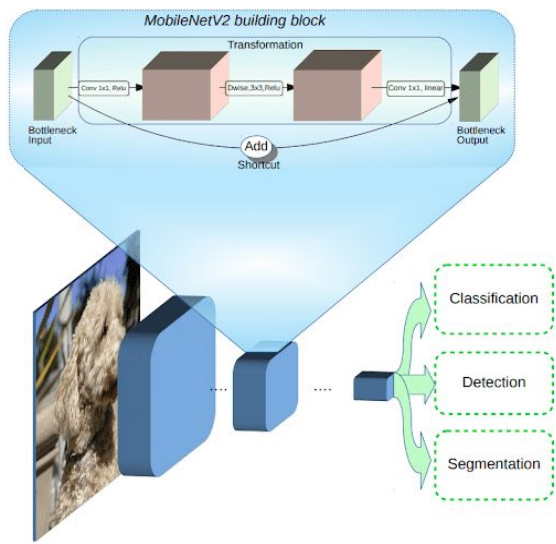

Overview of MobileNetV2 Architecture. Blue blocks represent composite convolutional building blocks as shown above.

The iOS application was built on Model-View-Controller (MVC) architecture, in which each of the components will handle a particular aspect of the application. This is one of the common architectures to build an iOS application. The View component is used to build all UIs for the application while the Model component takes charge of all the data logics. Finally, the controller will act as an interface to handle the business logics in Model and send output to View for rendering. The reason we chose this architecture is because it is very easy to prototype and develop applications.

## 3.3 Methods and techniques

We used various methods and techniques to implement the models and the application. First of all, the input images are resized to (224, 224) to be able to fit into our models. These data are also ensured to be converted into RGB channels before normalizing data input into range [0,1] for numerical stability purpose.

The second method we used is data augmentation. This is a technique to increase the number of images in the dataset, hence hopefully increase models' performance, by applying image transformation such as image flipping, image rotation or image center cropping. In Tensorflow, data augmentations are defined beforehand and the real data augmentation will happen during the training.process.

Another method we mentioned multiple times is transfer learning. This is a method in which we used a developed model for a particular task, in this case is MobileNetV2, as a starting point for our model. This so-called pre-trained model contains previous knowledge of learning images classification just like VGG 16 does. With this method, we want to utilize this knowledge from MobileNetV2 to solve our related problem of classifying waste images. There are many approaches in the transfer learning method that are suitable for different situations. For our case, because the dataset is small and our task is very similar to the task MobileV2 had done, we use this model as a base for low-level extractor and build a shallow model upon them. In other words, all of the heavy computation for feature engineering has been completed by the MobileNetV2 model and we only need to modify it so that it can learn the high-level features.

During the training process, we also employ early stopping technique to avoid overtraining. To explain it, there is a compromise between too much training and too little training. Too little training, the model will not fully understand the dataset and hence underfit on the test set while too much training will cause an overfit problem on the test set. Therefore, it is very important to stop at the right epoch where the generalization error tends to behave badly. The idea here is that we will stop training the model when either the validation error starts to increase or the validation accuracy starts to decrease in three continuous epochs. Moreover, we will add a callback method so that the best model during training will be saved for later use.

To further improve the performance of our model on mobile devices, post-training quantization is another technique that was used in our project. As we know, mobile devices are smaller than a computer and have limited resources. Therefore, such a normal Tensorflow model in HDFS format will not perform as it does on mobile devices. Hence, the method of quantization will help us reduce the model size without too much delegation in model accuracy and still can improve the latency of CPU or hardware acceleration. We used the simplest form of quantization, dynamic range quantization, to quantize parameters to float 16 while the model executes with float32 operation. Before the inference part, the parameters' weight are converted from floating point to 8-bit precision integer. Coming to inference time, it will be converted back to floating point. This process is only conducted once and also cached in the memory for the purpose of reducing latency.

Besides the methods mentioned above, we also try to use other techniques to improve the model's performance.

-Batch Normalization:

The batch normalization can solve internal covariate problems. The internal covariate will cause the difference of input distribution and output distribution in each layer. It will affect our fully connected layer to get complex distribution and the softmax() is hard to calculate final result (totally wrong result). We add some batch normalization into our layer distribution to keep each layer with the same distribution of input and output. Therefore, it enhances the prediction accuracy.

-Drop out:

During the training time, our model can easily drop into overfitting. In addition, our model has complex layer distribution with too many cells. This will cause our model to have poor performance in the test set. The drop rate is used to ignore parts of cells in some layer which is to reduce our model complexity. Therefore, in our layer distribution, we have this technique to reduce numbers of cells and calculation load during training time of our model.

-Use momentum:

Our model is using the loss function to update its weight. However, when the loss function curve is plain, our model is hard to learn and update its weight. Using Momentum is helpful for our model to update its weight easily.

-Adjust leaning rate:

Because our model has too many layers, each time to update its corresponding weight, the proper learning rate will decide whether the gradient will be exploding or vanishing. When the learning rate is large, the gradient will have exponential growth and our weight will tend to be infinite. If our learning rate is too small, it also causes our model learning hardly and the weight will not update. We have tried some many values of learning rate, like 1, 0.1, 0.01, 0.001 etc. In the end, we found that when the learning rate is equal to 0.001, our model will have a good learning state and get best performance in the test set.

# 4. Project Management

## 4.1. Brief intro to Project Management

Due to the restriction of COVID-19, it is obvious that it stops the face-to-face meeting between me and my team members and negatively influenced our performance on the project, whatever, we still managed our tasks well by using different online communication and management tools, for example, Monday.com for managing tasks, Google collab for code and test purpose, and Zoom for required team meetings. Fortunately, we 3 teammates are all in Melbourne so there are no internet delays or differences in time zones.

## 4.2. What approach to project management was used. If a standard approach was adapted, how and why were modification made

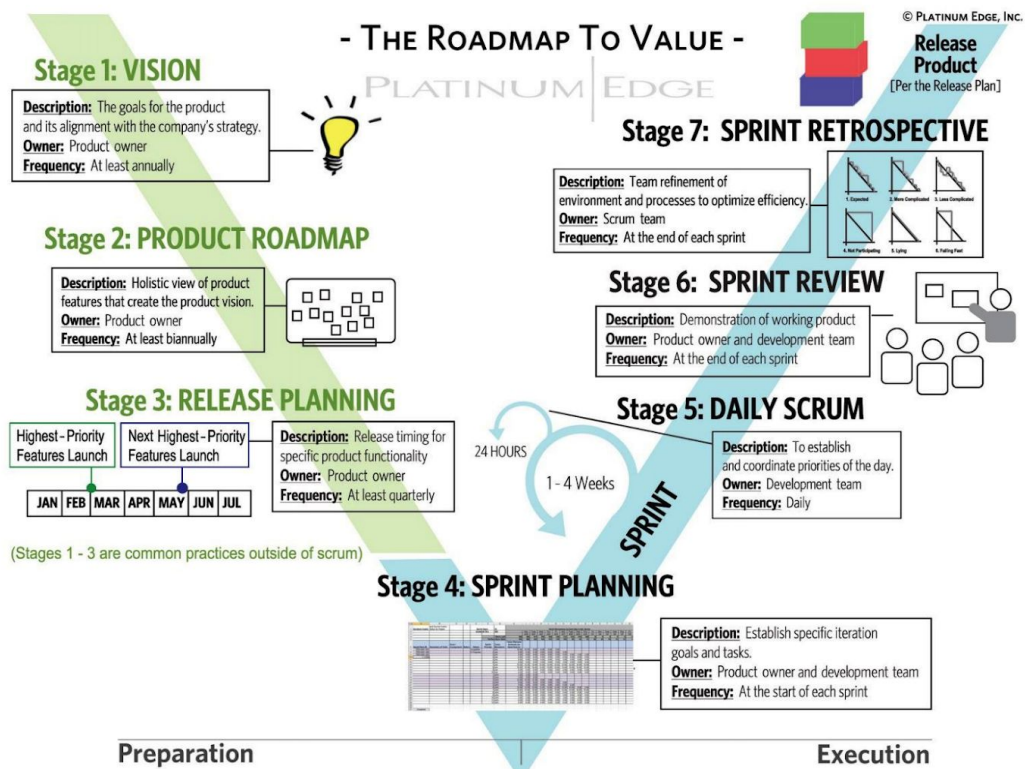

For our team, the agile management approaches were used. The word 'agile' means flexible, the project would become flexible if a software development team uses this kind of project management approach, means that the project team allows it to adapt and able to respond to changes at any time. It is useful for a project team if the project needs to frequently change its requirement and has unstable deliverables. The agile management approach has some useful characteristics, such as dynamic requirements, the major activities can repeat until the results get corrected, it can deliver small deliverables frequently, and the goal is to get customer value via frequent deliveries and feedback.

Moreover, Agile is so good when compared with the traditional software development life cycle. There is an agile roadmap, stage 1 to stage 7, Vision, Product roadmap, Release planning, Sprint Planning, Daily Scrum, Sprint view, Sprint retrospective.

1. The product vision is a definition of what the product is, how it will support the company or organization's strategy, who will use the product, and why people will use the product.
2. The product roadmap is a high-level view of the product requirements, with a loose time frame for when it is required (It is expected to update your product roadmap throughout the project).
3. Release planning conducts 2 activities, revisiting the product backlog and deciding the goal and the deadline. (Agile approach will have many releases and the one with the highest priority will be done first.)
4. Sprint planning is used to decide how long the sprint is, one project can have several sprints and work should be done within the sprints. (Sprints should last 1, 2, 3 and 4 weeks, and the maximum length should not be longer than 4 weeks.)
5. During each sprint, the development team should have a daily scrum meeting to discuss what has been done yesterday and what is the goal today and address the issues that may occur.
6. Sprint view, it is a meeting used to review and demonstrate the user stories completed by the project team in the sprint. The stakeholders of the project are able to review the product and give feedback to the team.
7. Sprint retrospective. It is held at the end of each sprint and it is a meeting where the scrum master, product owner and the development team discuss how the sprint went and what improvement they can make to the product. (The goal of sprint retrospective is to continuously improve the processes.

## 4.3. Execution

As described above, our project team uses agile project management approach and this approach executed successfully, agile is quite a suitable approach for a team with 3 undergraduate students because its flexibility counters our dynamic changes in planning.

For the 1st part of the approach is the stage one, Defining the product vision. 3 of us are first time working in the project developing environment and we were not sure what we are going to create, whether we had a lot of research on waste management last semester. Every decision is not fixed and we might change our decision when it demands, so firstly we decided to build a website that allows users to upload the photos of rubbish and further the web application will define the classification of the rubbish in the photo, the user of the application is everyone who has no idea on which bin is going to throw to, that is a web application because anyone can access it via mobile device or personal pc. This decision was inherited from last semester so we used it as the starting point and may be changed in the future.

In stage two, we are going to create a product roadmap to define the high-level view of the requirements. We made some changes to the product vision from stage one, we decided to create an IOS application after a few weeks meeting because we think that we are more familiar with IOS development. The product roadmap should contain 3 main parts, building CNN, apply transfer learning and IOS development. The roadmap is built based on the requirements of the product.(Requirements table is attach in the appendix)

| Building model with CNN | who | Subite… | Status | Due date | Priority |
|---|---|---|---|---|---|
| Build the CNN model | | | Done | | |
| Apply it to dataset | | | Done | | |
| Testing and evaluate performance | | | Done | ✓ Sep 20 | High |

| Building model with Transfer learning | who | Subite… | Status | Due date | Priority |
|---|---|---|---|---|---|
| Find a model for transfer learning | L | | Done | | |
| Augment the model to suitable with current dat… | L | | Done | | |
| Apply it to dataset | L | | Done | | |
| Testing and evaluate performance | L | | Done | ✓ Oct 4 | Medium |

| Deploy to mobile app | who | Subite… | Status | Due date | Priority |
|---|---|---|---|---|---|
| Reasearch on Tensorflow Lite or other method … | | | Done | | High |
| Establish pipeline to be ready to deploy once m… | | | Done | | Medium |
| Deploy and configure it | | | Done | ✓ Oct 17 | Low |

We step into stage 3 in week6, we finally set up our product goal and the deadline for each task. We will first build our own deep learning model which is Convolutional

Neural Network for our dataset and then use transfer learning to build another model, finally compare the performance of both machine learning models and use the best one to create the IOS application. The time limit is also assigned for each task, the deadline for building our own CNN is before the mid-term break(Sep 20$^{th}$), transfer learning must be done at the end of mid-term break(Oct 4$^{th}$), and the mobile app development should finish in week9(Oct 17$^{th}$) because we have code demonstration in Week10 workshop.

In later week6, we are moving into stage 4 and it is the most important stage for software development as the main execution steps are in this stage. We have three 2-week sprints in our project schedule. For the planning part for each sprint, we analysed the resource we gathered, for instance, some examples of CNN networks we learnt, great pre-trained models for transfer learning and online tutorials to get ideas. During the development of our own CNN, we had similar experience on developing models in the unit of FIT3181, and luckily, we are all taking that unit in this semester, therefore, we worked on this task together to find out the best model that has highest accuracy. After successfully building our own CNN model, we moved to the next developing sprint during the mid term break. Due to the high amount of assignments on midterm break, we decided to speed up our progress and one of us will start looking at the IOS development, another person will focus on the transfer learning part, the last one will start testing on the CNN model we built.

The stage five, six and seven are conducted in all 3 sprints. We do not have any meeting in between each sprint unless there is a serious issue raised or an important decision needs to make. During the sprints, we let each person continuously update their progress via Monday. We had sprint view and sprint retrospective in each sprint, but they did not take a large part of the group meeting, because we knew each person's tasks quite well and we can solve the issue as soon as possible, additional improvement we can make is limited because we do not have too much time to spend on developing the software because we are quite busy near the end of the term.

d. Project resources, execution, management and planning
Project resources
Project resources are what we depend on to get work done, they are different which depend on which type of the project is. For our project on the topic of waste management, the project resources are related to IT architecture, for example, physical equipment, software, communication techniques, development methodologies and modelling tools. Resources do matter for a project team because we must ensure all resources are available during the project execution period, we take risks if the resources are unavailable when we try to use the resources because it may halt the project or let the progress behind the schedule.

Firstly, the physical equipment for our project is quite simple to get. We all have laptops for programming purposes, and this required resource is also listed in the requirement table. The software and communication techniques are all free to use for us, such as the free GPU on google collab that increases the efficiency of machine learning, we use the free zoom account that is provided by University, also the free social media software Facebook for communication. The development methodologies and modelling tools are learnt from other units, all of us are taking the unit FIT3181 this semester, so it is possible to learn the main technique of machine learning by seeking help from the unit tutor. For example, we use the modelling tool Tensorflow, it is a powerful python machine learning library and free to use.

Management
The management is mainly handled by using Monday.com, a powerful task management tool. Although we do not have meetings between each sprint unless there are important discussions needed to conduct. We strongly advise teammates to update their task status on Monday, so all of us can see others' current progress. The communication process in our group was significantly improved as all members became proactive and responded very quickly. Monday.com also sends notifications when there is an update to github version control.

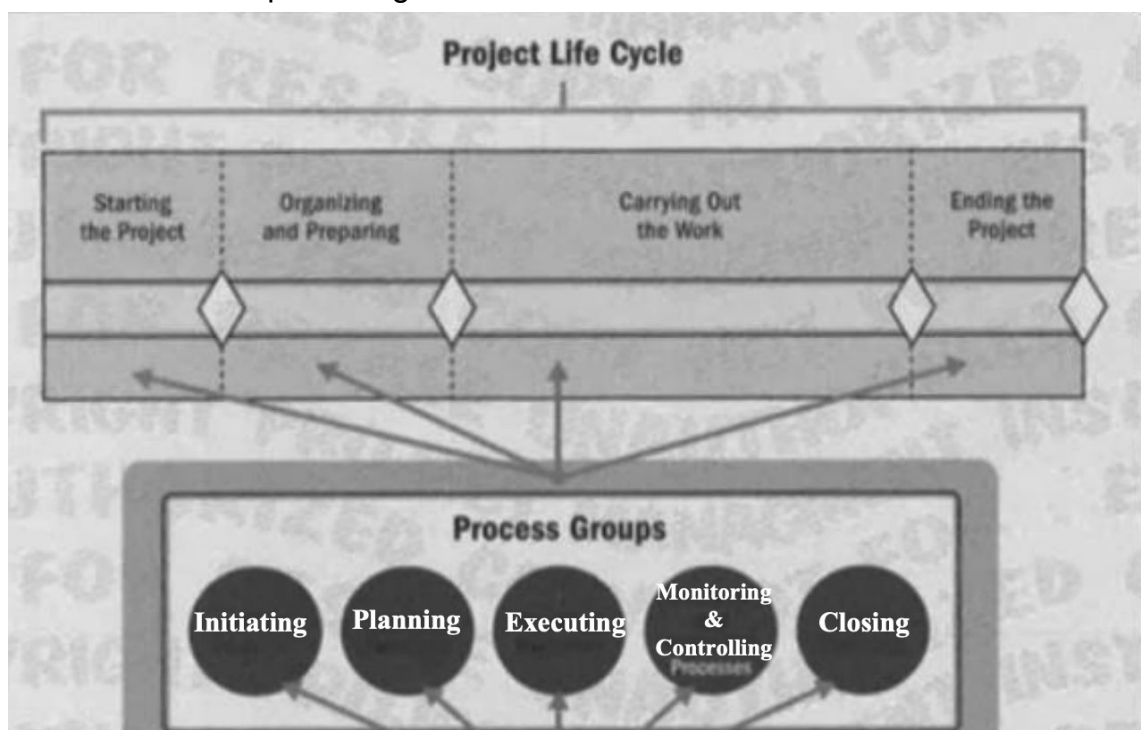

Referring to the picture above, Planning and execution are the phase 2 and 3 in the project life cycle.

Planning & Executing
The purpose of the planning phase is to lay down a detailed strategy of how the project has to perform and how to get successful. In our project, we spent about first 5 weeks to make the plan for our project, we decided what implementation we are going to use and assigned tasks to everyone, the planning phase is a little bit long I think because we do not have enough knowledge to practical start the project, that is the reason why we spent more time on doing research and decided what exactly we are going to make. To let the project succeed, we firstly planned to build a website application, but we changed our plan later because we are more familiar with IOS development. We also decided that we are going to use the Agile approach to manage our execution phase. We have 2-week sprints for developing our products.

In the execution phase, the phase started at the middle of the term, we have three 2-week sprints to build 3 different things. In this phase, the decision and activities defined in the planning phase are implemented, we firstly started the development of our own CNN, secondly the transfer learning model and lastly the IOS application. The phase 4 Monitoring and controlling is also combined into the execution phase because we also monitor and control the progress of execution every day and we keep checking everyone's progress. We use the online programming tool Google collab so everyone can see the error message online and fix it.

## 4.4. Risk management

Risk management is the process of identifying, analysing and responding to risk throughout the life of a project. If a team can handle any problem easily when responding to a risk identified, it means that the team must have well-prepared resources and strong ability to tackle potential risks.

We believed that our project team did well on risk management because we identified the potential risks at the planning stage, and we have enough resources to use when risks occur.
- Identify the risks by brainstorming and interviewing
Brainstorming is a famous technique especially for a group of people, we believe that brainstorming could be one of the best techniques to identify the risks. I remember that we ran a brainstorming session at the planning stage, once one of us proposed that we can do image classification using machine learning technique, we will then generate some potential shortage, for example, lack knowledge on image manipulation, no related development experience.
- Use a risk register table to manage the risks identified
A risk register table is the document that contains the results of various risks, also a tool for documenting potential risks events and related data. Our team members always keep one copy in hand and read it before they start executing in order to prevent unnecessary trouble.

The image below is a copy of the risk register table.

| No. | Rank | Risk | Description | Category | Root Cause | Triggers | Potential Responses | Risk Owner | Probability | Impact | Status |
|---|---|---|---|---|---|---|---|---|---|---|---|
| 1 | 3 | lack of data amount | We cannot get enough data to do the approach of machine learning. | Internal-technical | The testing result does not meet our expectation. | The testing result does not meet our expectation | Find more image data we can collect and let the model train on the new data. | The person who is in charge to build the classification model. | 2 | 8 | 16 |
| 2 | 2 | lack of technical skills | Our team have no idea how to build a IOS application. | Internal-team management | Our team have no knowledge on building a IOS app. | When we are stepping into the final application coding part, no one can coding for website. | Learn to build the IOS app from the internet or interview the technical man who is familiar with mobile-end development. | The leader in the team as he should let us know the related study area. | 1 | 7 | 7 |
| 3 | 6 | bad team communication | no-good contacting between team members. | internal-management | Leader of the team does not organize the time well. | Always a few people cannot attend weekly meetings. | Let the leader beaware on time management and make sure everyone can attend the meeting on the schedule. | Leader. | 1 | 5 | 5 |
| 4 | 1 | lack of technical skills | The implementation of machine learning is always getting wrong. | internal-technical | We have lack of knowledge on machine learning technique. | We have no idea how to build the machine learning model when we are going to move to the ML part. | Try to get help from tutor or the people has similar project experience. Self-learning from internet is also acceptable. | Technical man in the team who takes the responsibility of building the model. | 3 | 7 | 21 |
| 5 | 4 | bad UI design | The UI design on IOS end is to complicated and hard to use. | internal-technical | Lack of knowledge on design an easy-using UI. | Does not pass the user acceptance test. | Learn and view the great UI design on the internet, learn design from the good ones. | The UI designer in the team should take responsibility. | 4 | 3 | 12 |
| 6 | 5 | low quality of images | The quality of images uploaded by users is not high enough. | internal-technical | Not declearation of the quality or format of the images. | The classification model does not work well on images with low quality. | Define the quality of the image and show the statement to users. | The person who tests on the performance should take care of it. | 1 | 4 | 4 |

## 4.5. Limitations encountered during project management with discussion

Limitations:
- Covid-19

Covid-19 is a most serious element this year, it influences our team's performance a lot. It stops face-to-face meetings between project team members and limits the communication quality for our team. If there is no Covid-19, ideally, we can meet on campus every day and form more cohesive communication with each other. Whatever, we still processed to the last stage of report writing, everything's beginning is hard, but we can make it easier if we try hard on our objectives, whether who am I grouping with, we will be familiar to each other after passing a long period of time.

- Study pressure

We all 3 people are final year students, so we have graduation pressure at the end of the semester, we were busy with the study of other units. One of us has 5 units this semester, the rest of us have 3 units, which results in differences in time organizing and further our meeting time becomes shorter and shorter. At the end of IOS application development, we chose not to develop more functions on our product because we do not have enough time to spend on this unit and we have to take care of other units. However, our product still meets the requirement established and can classify rubbish quite well.

- Not initiative to communicate

I think at the beginning of the semester, we are not familiar with each other because we have not seen each other before or never face-to-face met. Therefore, our communication was not active and we did not talk much during the meeting. But this is not a big deal, every relationship starts with hardship, we believe that we will be familiar with each other if we spend more time together.

# 5. Outcomes: this is an important part of the report.

### 5.1. Results achieved/product delivered

Our product is successfully deployed on iPhone 8 Plus. The application can capture objects put before the camera continuously and transfer it to the backend. The application backend will do inference and produce the result accordingly. The result is then sent back and displayed on the view of an iPhone. The UI for modifying the number of kernels also works as we planned. The app functionally works as we presented in our demo presentation.

Besides that, we also completed a code/test report in which we include all source code coming with explanations and list of tests we conducted. Plus, there is a user guideline as well as technical guideline so that other people can easily understand and preproduct our project

### 5.2. How are requirements met

For this project, building model requirements are all achieved in a complete manner. As stated above, we created a model that is compatible with mobile devices and also deployed it successfully. For the mobile application, we met 80% percent of our requirements. The 20% remaining requirements are about developing a more complete UI for the app that will show the classified trash statistics. We did not complete those requirements mostly due to limited skills in iOS development in general.

### 5.3. Justification of decisions made

During this project, our team made multiple decisions to the project overall. They are switching the project from object detection problem to image classification problem; using transfer learning to build a powerful model with MobileNetV2; use iOS as the main platform for development

Firstly, we will discuss the decision of switching from object detection to image classification. Earlier in the project, we wanted to build a dynamic object detection in real time application with Tensorflow and OpenCV. We tried to build the model with the TrashNet dataset but the results we got were much lower than we expected. We realized that these poor results were due to the limited number of images in the dataset. It is also possible that we implemented it wrong. As we were in week 2 without any progress, we decided to switch back to image classification because we found a way that makes our app classify images continuously. Moreover, as we were consulted by a tutor, we found that there is no right or wrong method to implement this project topic but project deliverables are more important. Therefore, we finally came to the decision to build an image classification model.

The second decision is using a transfer learning algorithm to improve our model. A scratch model only achieved an accuracy of around 70%, which is not as good as we expected. As we did research on different techniques to leverage our model, we came across transfer learning technique, which is very easy to implement in Tensorflow thanks to its robustness. Therefore, we tried to implement a model with MobileNetV2 as this model is built to be more efficient on mobile devices than other devices. In the result, we trained a model with around 85% validation accuracy and 82% testing accuracy.

At first, we wanted to build cross platform applications on both iOS and Android using Flutter or React Native. However, due to COVID-19, I didn't have access to an Android phone. Moreover, there is limited documentation and support found from Google for Flutter. Therefore, we think it is better that we should focus on iOS platform in order to produce some deliverables on time.

During the making of this project,

## 5.4. Discussion of all results

The accuracy we received for the first customized model is around 66 ~ 68% for 50 epochs. The second model built upon VGG16 architecture is slightly better than the previous model in terms of accuracy with 68% ~ 70% for 30 epochs but it is painfully slow to train with more than 138 million parameters. Another downside of this network is that it is a large network of over 533MB. Therefore, the deploying task was tiresome to us. On the other hand, the model using transfer learning method achieved a highest validation accuracy of more than 85% with much faster training speed. With a smaller size and optimized architecture for mobile devices, we chose this model to be the model for our application and converted it to Tensorflow Lite model. Moreover, with post-training quatization technique, we found that our final model is around 2 times smaller and 2 times lower in the inference time than the model without quantization.

In terms of iOS application, we tested the app with unit tests to ensure that the functionalities of each class work normally without bugs. We also did the performance test to assess how it interacts in the production environment. We found that the app works well when there is a plain background behind while behaving poorly when there are other objects in the background. Therefore, we will consider to improve it further by removing background technique or adding noise to the background so that the model can learn images more robustly. Moreover, we found that for this task, one thread is good enough for the model to run. When we increase the number of threads, the inference time also increases accordingly.

### 5.5. Limitations of project outcomes

There are few limitations in our project. Firstly, the application is quite simple as it doesn't have many UIs. It cannot be put into production immediately due to its simplicity. We wanted to develop more features to make it more complete. However, due to lack of familiarity with Swift and iOS development in general, we couldn't reach the production-level point.

The second limitation is the MVC architecture of our application. Although this architecture is encouraged by Apple, the robust connection of View and Controller makes it hard to do full complete testing with unit tests. We cannot easily isolate each layer as they depend upon each other. If we add more functions without proper testing, the application can easily break as a result. Therefore, in order to test our app, we have to design a workaround solution by display the results like inference time to the view in order to measure the performance of the model

The third limitation of the project is that we couldn't find a large enough dataset to make our model more robust. Currently, it can only predict 6 classes with very poor performance in product-level conditions because the model cannot recognize well objects with too much distraction in the background. Therefore, in order to leverage this project, it requires us to find or build a larger dataset with more variation so that the model can generalize the dataset better.

## 5.6 Discussion of improvements and possible future works

First of all, we want to expand the dataset of trash. Trash without proper processing is a very problematic issue nowadays. There should be a larger dataset so that we can utilize deep learning advances to tackle this environmental problem. However, building the dataset on our own is not possible as the model needs hundreds of images of each class to classify the dataset well. Therefore, we think that we can use the power of community to solve this problem. In particular, we want to create an open source project where community members can capture images from their local and upload it into a cloud dataset and label it. With the solution, we hope to gather millions of images within 1 or 2 years.

The second improvement is adding more features to our application. Firstly, we want to complete 20% of our listed requirements at the beginning of the semester. After that, we will consider adding more features such as snapshot a picture and classify it. In addition, we want to combine the app with a smart bin to make it a complete solution to many households.

Another issue we want to further improve is migrating the architecture of MVC to a more robust architecture like VIPER, a short term for View, Interator, Presenter, Entity and Routing. Upon our research, this is a dynamic architecture to build iOS applications where the app's logical is divided into many distinct layers. This architecture helps us isolate each layer dependency more easily. However, we need to do further research on the topic as such dynamic architecture might not be good for a small application as it requires much more effort to build it successfully.

## 6. Critical discussion and Conclusion

In general, the project outcome is positive since we finally delivered a workable product. However, we are not totally satisfied as we didn't meet all of the requirements we planned earlier. We learnt that team management and product management are critical to the success of such a project. We saw the effect of not having everybody on the same page as well as learnt how to effectively write solutions under a restricted time frame which involved a constant need for communication. The process of making this project also highlights how important interpersonal skills are in developing a sophisticated application

Compared with our original proposal, there have been many changes in our project. In the proposal, we expected that we could complete the software by week 8 or 9 so that we can focus on writing the final report. However, we were ambitious and wanted to develop a mega project but didn't identify all the requirements and risks carefully enough. Hence, we decided to go back to our original project proposal but deploy on a mobile device instead of building a web application. We also learnt how to manage a team under agile methodology. This framework allows us to build and deploy products faster, hence creating the motivations for us to continue to develop and refactor app's functionality. If we used an iterative approach, we think that it is very hard to keep every member accountable as they don't see such a product is being rolled out.

Agile approach also allows us to be flexible with our decision making at the critical time. This is one of the benefits that we found helpful when we work with this framework. Waterfall method has its benefits. However, when we look back on the whole process of making this project, we found that trials and errors would be more useful to us. If our topic is not waste management at the very beginning, we thought waterfall or agile both workable. For an open topic with many freedoms like our topic, we feel satisfied with our framework pick at the early stages. Otherwise, we could hardly deliver any products at this stage.

In conclusion, our project utilized a transfer learning technique to build a powerful model using MobileNetV2 as backbone with 85% validation accuracy that is compatible with low latency devices. This model is then deployed in an iOS application designed with MVC architecture. The application, however, is not ready for production as it performs poorly on images with background noise during the testing process. To further improve it, we possibly create an open source project that aims to expand the TrashNet dataset, develop more features for current app as well as seek to combine it with smart bin so that users can benefit the most from an IoT ecosystem.

# 7. Appendix (if applicable)

- Requirements table

| ID | Requirements (Functional or Non-Functional) | Assumption(s) and/or Customer Need(s) | Category | Source | Status |
|---|---|---|---|---|---|
| 001 | Non-Functional | Laptops with internet access for 3 team members. | hardware | The demand of the project team, the team needs the devices to start the project. All of 3 team members have our own laptops. | Completed |
| 002 | Non-Functional | We require thousands, or huge number of image data that can be used on our data training. | Software | This training dataset can be found from the open data source, such as kaggle. | Completed |
| 003 | Functional | machine learning technique | Software | This can be done by all of us. We have ability to achieve it. | Completed |
| 004 | Functional | data visualisation as the final product | Software | The can be done by team's data analyst. | Completed |
| 005 | functional | Develop a website that show to users and upload images to the server. | software | One of our teammember will take of the the website design. | Completed |
| 006 | Functional | The technique use to convert image into numerical form, and let the computer read it to achieve data modeling. | Software | We will do more research on how it exactly works and make the implementation on it. We can make inteview on the people who has similar project experience. | Completed |
| 007 | Non-functional | We need scirpt editing software, such as Pycharm, R studio and more front end languages editor. | software | Download the related software/tools from the internet. | Completed |
| 008 | Functional | Producing great UI design for the users, guarantee that the design is suitable for all people, easy to use by following the instructions. | sofware | We have technical man who is familiar with UI design, one of our team member will get the UI design for the IOS application. | Completed |
| 009 | Functional | Showing the classification of the rubbish inputed online. | Software | This function will be handled by our technical person, it requires the guy who is pro in front-end programming. | Completed |

- Task allocation in report writing

| Name | Section |
|---|---|
| Fei | Background, Methodology and Literature Review |
| Liqin | Introduction and Project management |
| An | Methodology, Outcomes and Conclusion |

# 8. References: References need to be provided in a correct and consistent format (APA)